\documentclass[fleqn,usenatbib]{mnras}

\usepackage{newtxtext,newtxmath}

\usepackage[T1]{fontenc}

\DeclareRobustCommand{\VAN}[3]{#2}
\let\VANthebibliography\thebibliography
\def\thebibliography{\DeclareRobustCommand{\VAN}[3]{##3}\VANthebibliography}

\usepackage{graphicx}	
\usepackage{amsmath}	
\usepackage{amssymb}	
\usepackage{gensymb}				
\usepackage{subcaption}				
\usepackage[acronym]{glossaries}	
\setacronymstyle{long-short}

\usepackage{xcolor} 				
\usepackage{hyperref} 				

\newcommand{\lmin}{$\lambda_{\rm min}$} 
\newcommand{\lmax}{$\lambda_{\rm max}$} 
\newcommand{\numax}{$\nu_{\rm max}$} 	
\newcommand{\bmin}{$B_{\rm min}$}		
\newcommand{\SR}{$r_{02}$} 				
\newcommand{\msun}{$\rm M_{\odot}$} 	
\newcommand{\rsun}{$\rm R_{\odot}$} 	
\newcommand{\Brel}{$B_{\rm rel}$} 		
\newcommand{\Kepler}{\emph{Kepler}} 	
\newcommand{\tess}{\emph{TESS}} 	
\newcommand{\el}[1]{$l\!=\!#1$}			
\newcommand{\shift}[1]{$\delta\nu_{#1}$}

\newacronym{ssm}{SSM}{Standard Solar Model}
\newacronym{mcmc}{MCMC}{Markov Chain Monte Carlo}
\newacronym{AIMS}{AIMS}{Asteroseismic Inference on a Massive Scale}
\newacronym{YGM}{YGM}{Yale Grid-based Modelling}

\title[Magnetic activity asteroseismic impact]{Impact of magnetic activity on inferred stellar properties of main sequence Sun-like stars}

\author[A. E. L. Thomas et al.]{
Alexandra E. L. Thomas$^{1,2}$\thanks{E-mail: axt367@student.bham.ac.uk (AELT)},
William J. Chaplin$^{1,2}$,
Sarbani Basu $^{3}$,
\newauthor{Ben Rendle $^{1,2}$,}
Guy Davies $^{1,2}$,
Andrea Miglio $^{1,2}$
\\
$^{1}$School of Physics \& Astronomy, University of Birmingham, Edgbaston, Birmingham B15 2TT, UK\\
$^{2}$Stellar Astrophysics Centre, Aarhus University, Ny Munkegade 120, DK-8000 Aarhus C, Denmark\\
$^{3}$Department of Astronomy, Yale University, PO Box 208101, New Haven, CT 06520-8101, USA
}

\date{Accepted 2021 February 4. Received 2021 January 11; in original form 2020 October 30}

\pubyear{2021}

\begin{document}
\label{firstpage}
\pagerange{\pageref{firstpage}--\pageref{lastpage}}
\maketitle

\begin{abstract}
The oscillation frequencies observed in Sun-like stars are susceptible to being shifted by magnetic activity effects. The measured shifts depend on a complex relationship involving the mode type, the field strength and spatial distribution of activity, as well as the inclination angle of the star. Evidence of these shifts is also present in frequency separation ratios which are often used when inferring global properties of stars in order to avoid surface effects. However, one assumption when using frequency ratios for this purpose is that there are no near-surface perturbations that are non-spherically symmetric. In this work, we studied the impact on inferred stellar properties when using frequency ratios that are influenced by non-homogeneous activity distributions. We generate several sets of artificial oscillation frequencies with various amounts of shift and determine stellar properties using two separate pipelines. We find that for asteroseismic observations of Sun-like targets we can expect magnetic activity to affect mode frequencies which will bias the results from stellar modelling analysis. Although for most stellar properties this offset should be small, typically less than 0.5\% in mass, estimates of age and central hydrogen content can have an error of up to 5\% and 3\% respectively. We expect a larger frequency shift and therefore larger bias for more active stars. We also warn that for stars with very high or low inclination angles, the response of modes to activity is more easily observable in the separation ratios and hence will incur a larger bias.
\end{abstract}




\section{Introduction}
\label{sect: introduction}

Sun-like stars with outer convective zones pulsate due to turbulent motion within these layers. These perturbations excite acoustic waves within the star producing a rich spectrum of modes of oscillation which can be studied via asteroseismology to investigate internal stellar physics and determine global properties. Thanks to the high-resolution photometric observations from CoRoT \citep{2006ESASP1306...33B}, \Kepler\ \citep{2010Sci...327..977B,2014PASP..126..398H}, and more recently \tess \citep{2015JATIS...1a4003R}, we have measured acoustic oscillations for thousands of stars.
	
Surface magnetic activity is known to affect acoustic modes, changing their frequencies, enlarging damping rates and suppressing amplitudes \citep[e.g.][]{2002RvMP...74.1073C,2007MNRAS.377...17C,2007MNRAS.379L..16M,2015Sci...350..423F,2019FrASS...6...52K}. Several studies have found evidence of acoustic modes being shifted in frequency, an effect that varies with the 11-year solar activity cycle \citep[e.g.][]{1985Natur.318..449W,1989A&A...224..253P,1990Natur.345..322E,2002ApJ...580.1172H,2007MNRAS.377...17C,2014SSRv..186..191B} as well as with a quasi-biennial period \citep[e.g.][]{2010ApJ...718L..19F,2012MNRAS.420.1405B,2012A&A...539A.135S}. Similar behaviour has also been found in other solar-type stars \citep[e.g.][]{2010Sci...329.1032G,2016A&A...589A.118S,2018ApJ...852...46K,2018ApJS..237...17S}. \cite{2020MNRAS.496.4593K} showed that mode frequencies from main sequence and subgiant stars are particularly sensitive to perturbations from magnetic activity. The sizes of frequency shifts tell us about activity related changes in the outer layers of stellar interiors enabling us to probe activity cycles \citep{2017A&A...598A..77K,2018ApJS..237...17S}, and relative sizes of shifts can reveal information about the surface activity distribution \citep{2018A&A...611A..84S,2019MNRAS.485.3857T}. It is important to account for activity related effects on asteroseismic measurements since they are often used in stellar modelling or to infer global stellar properties such as mass, age or radius \citep[see e.g.][]{2015sac..book..437C,2019FrASS...6...52K,2019FrASS...6...41P,2019ApJ...883...65S}. 

When modelling stars the description of the near-surface layers is generally incomplete causing a discrepancy between theory and observation. Different techniques have been developed to deal with this (see Section \ref{eqn: SR}), however, generally these assume spherical symmetry within these regions which is not true for magnetic activity. The aim of this work was to determine the impact a non-homogeneous spread of near-surface activity has on the estimates of fundamental stellar properties made by asteroseismic modelling pipelines and in what situations the effect must be considered. Since this is a first attempt to quantify the expected bias due to magnetic activity, we considered Sun-like stars since the activity patterns on the Sun are well known. We generated artificial observations for Sun-like stars with various near-surface field strengths and spatial distributions of surface activity. We studied the difference between estimations of stellar models fitted to shifted and non-shifted sets of frequency separation ratios. Comparisons made between these results would reveal the impact of activity. The justification for using separation ratios is explained in Section \ref{sect: separation ratios}. In Sections \ref{sect: theory, freq shifts} and \ref{sect: data} we outline the model used to calculate activity-induced frequency shifts and our process for generating sets of frequencies. Section \ref{sect: stellar models} contains a description of the two stellar modelling pipelines used to infer global properties. Our results are shown in Section \ref{sect: results} followed by discussion and conclusions.

\section{Use of separation ratios}
\label{sect: separation ratios}

Asteroseismology is a powerful tool to infer fundamental properties of solar-type stars. With long-timebase photometry from, for example, \Kepler\ and \tess it is possible to resolve individual modes of oscillation in stellar spectra giving us a window into the inner workings of stars. Asteroseismic modelling pipelines can be used to obtain precise properties of stars using inputs of mode frequencies along with complimentary non-seismic data, typically, but not limited to, effective temperature, metallicity, and luminosity derived using parallaxes. More robust and higher accuracy constraints on stellar properties, including mass, radius and age, are possible with the inclusion of individual mode frequencies, or combinations of frequencies, as opposed to solely using global seismic quantities \citep[e.g.][]{2011ApJ...730...63G,2014A&A...569A..21L,2014ApJS..214...27M,2016A&A...592A..14R,silva_aguirre_standing_2017}. Using a set of input physics and evolutionary codes stellar models are computed, either on the fly or to build a predefined grid, where each stellar model corresponds to a combination of properties for a model star. For each model pulsation codes are then used to predict theoretical oscillation frequencies. Theoretical observables, including frequencies and additional non-seismic data, are fit to actual observations to obtain the best matching model and the corresponding stellar properties.

The oscillations are acoustic modes where pressure perturbations drive standing waves within a main sequence star. Spherical harmonics are used to describe the appearance of these modes on a sphere with oscillations usually described by three numbers:the radial order, $n$, the angular degree, $l$, and the azimuthal order, $m$. Solar-like oscillators produce a spectrum of modes, the frequencies of which depend on the star's properties and internal structure. However, for solar-type stars observed by \Kepler\ and \tess it is only possible to measure modes with $l\leq3$ due to geometric cancellation for higher degrees. 
	
When modelling oscillation frequencies there is a known systematic difference between models and observations called the surface effect. This frequency bias is caused by the incomplete modelling of near-surface layers of stars, for example by using approximations such as mixing-length theory, or inadequate modelling of the interactions between oscillations and convection (for more description see \cite{2018ApJ...869....8B} and references therein). There are however several methods to allow for this effect. Correction terms have been included to account for the offset \citep{2008ApJ...683L.175K,2013MNRAS.435..242G,2014A&A...568A.123B,2015A&A...583A.112S}, or the use of an asteroseismic phase to parameterise the frequency dependent difference between model and observation \citep{2015A&A...574A..45R}. 

Another method is to use combinations of frequencies when fitting rather than the individual frequencies themselves. These are known as `separation ratios' and are useful since they are roughly independent of the structure of surface regions of stars thereby mitigating the impact of the surface effect \citep{2003A&A...411..215R}. For this work we use the \SR\ separation ratio defined as 
\begin{equation}
	\begin{aligned}
	r_{02}(n) &= \frac{d_{02}(n)}{\Delta\nu_1(n)},\\
	\textnormal{where} \qquad \qquad\\
	d_{02}(n) &= \nu_{n,0}-\nu_{n-1,2} \:, \\
	\Delta\nu_1(n) &= \nu_{n,1}-\nu_{n-1,1} \:.
	\end{aligned}
	\label{eqn: SR}
\end{equation}
Here $\nu_{n,l}$ is the frequency of a mode with radial order $n$ and angular degree $l$, $d_{02}(n)$ is the small frequency separation between \el0 and \el2 modes, and $\Delta\nu_1(n)$ is the large frequency separation for \el1 modes. The sensitivity of low-$l$ modes to near-surface conditions is independent of the degree. Therefore the small separation is already fairly insensitive to surface layers since it calculates the difference between two modes of very similar frequency which both propagate in the near-surface regions. This sensitivity is reduced even further when taking the ratio of small to large separations. \cite{2003A&A...411..215R} compared stellar models with the same interior structure but different surface layers to illustrate that frequency ratios are much less sensitive to the poorly-modelled outer layers' conditions. As a result, despite some loss of information when taking ratios, separation ratios can be used to isolate the effects of the deep stellar interior, a key focus for those determining ages and evolutionary states of stars. Due to this sensitivity to central conditions and being almost unaffected by surface regions \cite{2013ApJ...769..141S} argue that more reliable stellar properties can be obtain by using separation ratios rather than oscillation frequencies themselves. \cite{2018ApJ...869....8B} showed that, as long as the surface effect is somehow compensated for when inferring stellar properties from models, then the obtained results are robust.

Nevertheless, one underlying assumption when using frequency ratios is that there are no non-spherically symmetric near-surface perturbations \citep{2005MNRAS.356..671O} which would induce frequency shifts that depend on the degree of the mode. One example source would be a surface magnetic activity distribution such as that seen on the Sun. The $\Delta\nu_l$ are calculated from modes with the same combination of $m$ and $l$ therefore changes in $\Delta\nu_l$ due to activity will be negligible. However, $d_{02}$ uses frequencies with different combinations of $m$ and $l$ which occupy different spatial regions on a star and hence experience different size shifts in the presence of non-spherically symmetric magnetic activity. For this reason $d_{02}$, and hence $r_{02}$, will carry a signature of magnetic field changes; for example the solar cycle. \citet{2005ApJ...635L.105C} studied this effect using Sun-as-star observations from BiSON \citep{1996SoPh..168....1C} and measured variations in the ratios with solar activity level. They attribute this change to acoustic asphericity from surface activity and advised that care must be taken to account for biases when using ratios from long data sets. For this reason in this study we use separation ratios when fitting stellar models, avoiding the need for a surface term and we know that any remaining effect will be caused by non-spherically symmetric activity.


\section{Modelling activity-induced frequency shifts}
\label{sect: theory, freq shifts}
The shift in frequencies due to surface activity will depend not only on the star's magnetic field strength but also the spatial distribution of the activity on the stellar surface. What's more, our ability to observe these shifts will depend on the inclination angle of the star since this dictates the mode visibility and hence our ability to observe mode components of different $l$ and $m$. For this work we build on the model from \cite{2019MNRAS.485.3857T} but summarise the main principles here.

The response of each mode depends on the field strength in the region where the corresponding acoustic wave propagates so the impact of a non-homogeneous distribution of activity on a mode will depend on the mode's spatial distribution. Assuming that the frequency shifts are caused by a source in the near-surface regions of the star, the shift experienced by modes of different $m$ and $l$ is given by \citep{2000MNRAS.313..411M}:
\begin{equation}
	\delta\nu_{lm} \propto \: \Big(l+\frac{1}{2}\Big) \: \frac{(l-|m|)!}{(l+|m|)!}  \int\limits_{\theta_{\rm min}}^{\theta_{\rm max}}|P^{|m|}_{l}\:(\cos\theta)|^2 \: B(\theta) \: \sin\theta \: d\theta.
	\label{eqn: freq shift general}
\end{equation}

From this we can see that a mode's sensitivity to activity has a term that depends on the mode $m$ and $l$ whose spatial response is described by the associated Legendre polynomials, $P^{|m|}_{l}(\cos\theta)$. This is combined with the magnetic field strength, $B(\theta)$, which is a function of the distribution of activity. The above describes this distribution using $\theta$, the co-latitude on the sphere, however, for the rest of this paper we will use the latitude, $\lambda$, where $\theta=(\frac{\pi}{2} -\lambda)$. The arrangement of activity on the stellar surface therefore determines the relative magnitudes of shifts for modes of different $l$ and $m$.
To calculate these shifts we assume the same top-hat model as \cite{2019MNRAS.485.3857T}, defined such that the magnetic activity is distributed in each hemisphere as a band of uniform field strength, $B$, lying between latitudes \lmin\ and \lmax, i.e.
\begin{equation}
B(\theta=\frac{\pi}{2}-\lambda)=\begin{cases}
B, & \text{if } \lambda_{\rm min}\leq\lambda\leq\lambda_{\rm max},\\
0, & \text{otherwise}.
\end{cases}
\label{eqn: B-field}
\end{equation}
The southern hemisphere is assumed to be a reflection of the northern hemisphere, since the globally coherent modes have no sensitivity to differences between hemispheres.

Although theoretically we can describe the response of individual $m$ modes to activity, it is not always possible to isolate them in a frequency spectrum. Typically, stellar modelling pipelines use only one frequency per $n$ and $l$ mode as input so the frequencies of the $m$ components must be combined. The relative visibility of each azimuthal mode depends on the inclination angle of the star, $i$. By introducing a weighting factor, $\alpha(i)$, which is a function of the inclination angle, \cite{2019MNRAS.485.3857T} defined how the contributions combine to give the central frequency of an \el1 multiplet. For this work we used this along with an equivalent weighting for the \el2 multiplet so as to include the mode visibility dependence on $i$. 
Therefore the \textit{measured} frequency shift of the combined $l$ peak will depend not only on the magnitudes of the individual $m$ shifts but also on the inclination angle.

It is well known that activity-induced frequency shifts have a dependence on the radial order of the mode \citep{1990Natur.345..779L,1998MNRAS.300.1077C}. Higher frequency modes have shallower upper reflection points than their lower-frequency counterparts and are therefore more sensitive to the perturbations in the layers closer to the stellar surface. That they therefore show larger frequency shifts has been observed for the Sun and also other solar-like oscillators \citep{2011A&A...530A.127S,2016A&A...589A.118S,2017A&A...598A..77K,2018A&A...611A..84S}. To account for this in our artificial data we adjusted the shifts to have a Sun-like frequency dependence as represented by a polynomial relation in frequency \citep{1998MNRAS.300.1077C,2017MNRAS.464.4777H}.

Using the same frequency shift model, \cite{2019MNRAS.485.3857T} found that the active latitudes required to produce the observed solar shifts (for solar cycle 23) extend between \lmin$=3.3\degree$ and \lmax$=40.6\degree$. This describes the spread of significant, large-scale field on the solar surface. We therefore re-parameterise Equation \ref{eqn: B-field} to replace $B$ by a relative field strength $B_{\rm rel}$, normalised to unity for the Sun:
\begin{equation}
B(\theta=\frac{\pi}{2}-\lambda)=\begin{cases}
B_{\rm rel}, & \text{if } \lambda_{\rm min}\leq\lambda\leq\lambda_{\rm max},\\
0, & \text{otherwise}.
\end{cases}
\label{eqn: B-field reparam}
\end{equation}
To enable this we introduced a multiplicative calibration constant $C_{\delta\nu}$ into the frequency shift calculation, turning the proportionality in Equation \ref{eqn: freq shift general} to an equality giving
\begin{equation}
\delta\nu_{lm} = \:  C_{\delta \nu} \, B_{\rm rel} \: \Big(l+\frac{1}{2}\Big) \: \frac{(l-|m|)!}{(l+|m|)!}  \int\limits_{\theta_{\rm min}}^{\theta_{\rm max}}|P^{|m|}_{l}\:(\cos\theta)|^2 \: \sin\theta \: d\theta.
\label{eqn: freq shift used}
\end{equation}
The value of this calibration constant is fixed to $C_{\delta\nu} = 0.8$ in order to produce modelled shifts that match those from BiSON observations for \Brel$=1$, $i=90\degree$, and the above mentioned latitude parameters pertaining to the Sun.

Our model uses Equation \ref{eqn: freq shift used} to calculate the frequency shift of each $m$ mode. As described above we used the weighted contributions of the azimuthal components to calculate the central frequencies of the shifted \el1 and \el2 combined peaks. Finally all frequencies were corrected for the radial order dependence. We used this model to generate artificial activity-affected frequencies sets.

\begin{figure}
	\centering
	\includegraphics[width=0.48\textwidth]{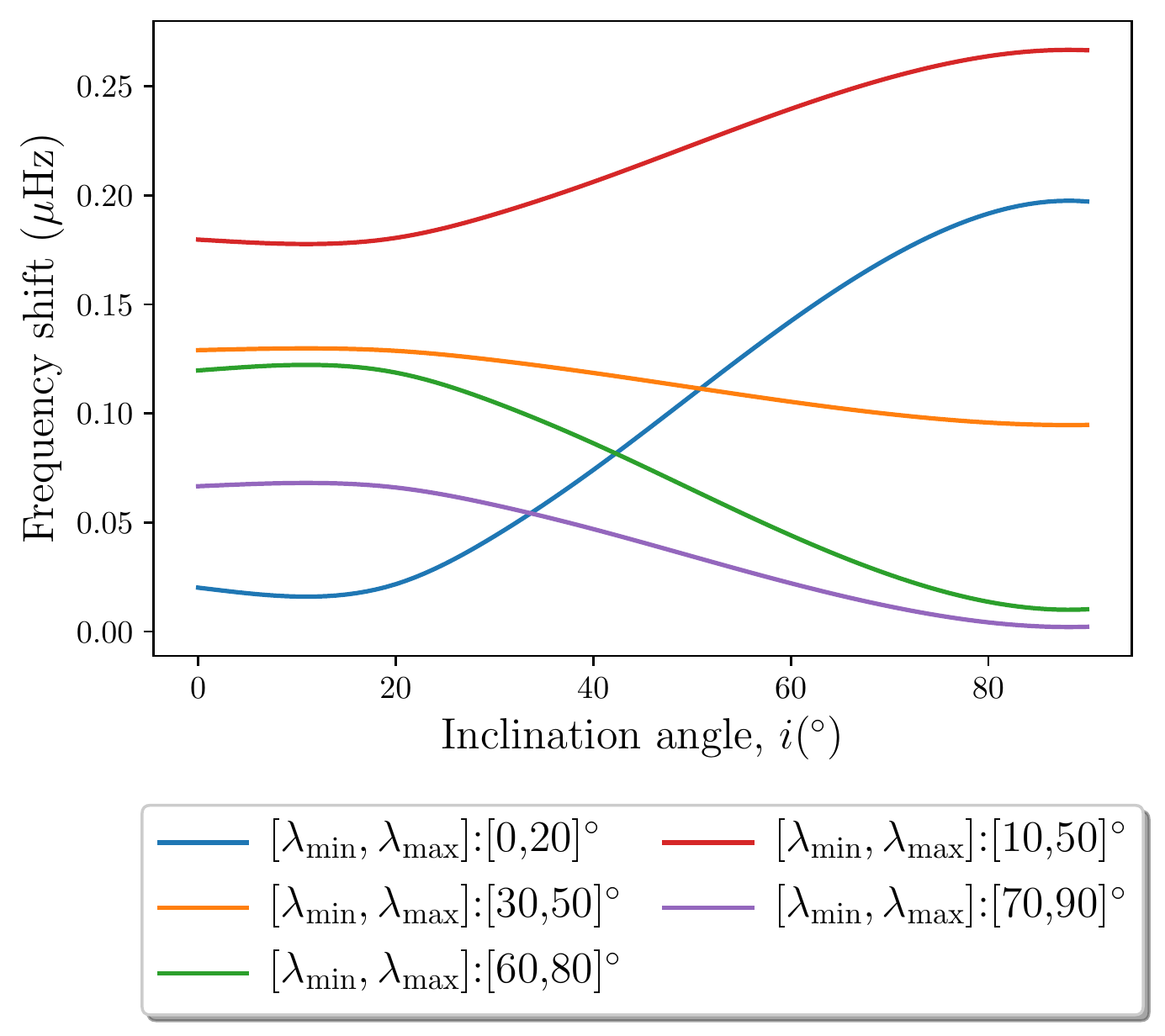}
	\caption{Example frequency shifts of \el1 modes at various inclination angles in response to different magnetic activity distributions. The field strength was kept constant.}
	\label{fig: example l=1 combinations}
\end{figure}

From the above we know that the measured (i.e. observed) frequency shift depends on the degree $l$, inclination angle $i$, relative magnetic field strength \Brel, and distribution of activity \lmin\ and \lmax. Increasing the magnetic field strength will induce a larger magnitude frequency shift. At certain inclination angles it will be easier to observe particular $m$ components due to the relative mode visibility. The response of modes and how we measure them is complex. Figure \ref{fig: example l=1 combinations} shows an example of how the measured frequency of a \el1 mode responds to changes in inclination angle for a few different \lmin\ and \lmax\ combinations. We can see that the response is not necessarily straightforward. The \el2 case is even less intuitive since there are five $m$ components to combine which each depend on the activity distribution and whose relative contribution to the measured \el2 frequency also depends on the inclination angle.

\section{Generating artificial frequency sets}
\label{sect: data}
We use two stellar modelling pipelines to infer properties for artificial stars from their `observed' frequencies: \gls{AIMS} and another grid-based approach which we call \gls{YGM} (see Section \ref{sect: stellar models} for more details). Both methods can fit to separation ratios thereby avoiding the need for a surface term. 

To build our artificial data sets we started with a set of `pristine' frequencies, free from any simulated magnetic activity effects. At an activity minimum we expect a more uniform distribution of activity across the stellar surface so all modes experience the same size of shift. This cancels out when taking the ratio of frequencies. Therefore by using ratios when fitting stellar models, our pristine separation ratios are equivalent to what we would observe on a field-free star, and also what we would expect to observe at minimum levels of stellar activity. \cite{2019MNRAS.489L..86C} showed that minimum-epoch solar p modes should have frequencies very close to field-free case. 

For the \gls{AIMS} analysis presented in Section \ref{sect: results} the pristine frequencies were taken from the model in the grid which was most similar to the Sun in terms of mass and age (4.61 Gyr). The pristine frequencies for the \gls{YGM} analysis were taken from a calibrated \gls{ssm}. This was created with the same input physics as the grid that was used for fitting, except for the atmospheric model, which was that of \citet{krisSwam_atmos} (see Section \ref{sect: stellar models} for details of the physics of the grid). As is usual in constructing \gls{ssm}s, we iterated over the mixing length parameter and the initial helium abundance in order to get a 1\msun\ model that has the correct radius and luminosity at the solar age (4.57 Gyr). The converged model has a mixing-length parameter of  2.1566 and an initial helium abundance of 0.2734. The model has a convection-zone helium abundance of 0.2447, and $Z/X$ of 0.02299. The base of the convection zone is at 0.71317\rsun. For completeness we repeated the analysis with the pristine frequency sets swapped; i.e. the \gls{AIMS} pipeline was also run with data sets based on the \gls{ssm} frequencies, and the \gls{YGM} analysis using frequency sets based on those from the most solar-like \gls{AIMS} model. The results were in agreement whichever set of pristine frequencies were used as a base.

Data sets were comprised of the 10 overtones of degrees \el0,1,2 centred on \numax\ (to match the procedure of \citealt{2014A&A...568A.123B}). Frequency uncertainties were taken from BiSON 1-year data and are comparable to uncertainties given by \Kepler\ data of duration a year or more from high-quality SNR targets.

Using the pristine data as the base, activity-affected frequency sets were generated by shifting the pristine modes according to our model and the chosen combination of \Brel, $i$, \lmin\ and \lmax. The artificial data were created to represent solar-like oscillators at various inclination angles and with a variety of magnetic activity strengths and distributions. 

\begin{figure}
	\centering
	\includegraphics[width=0.48\textwidth]{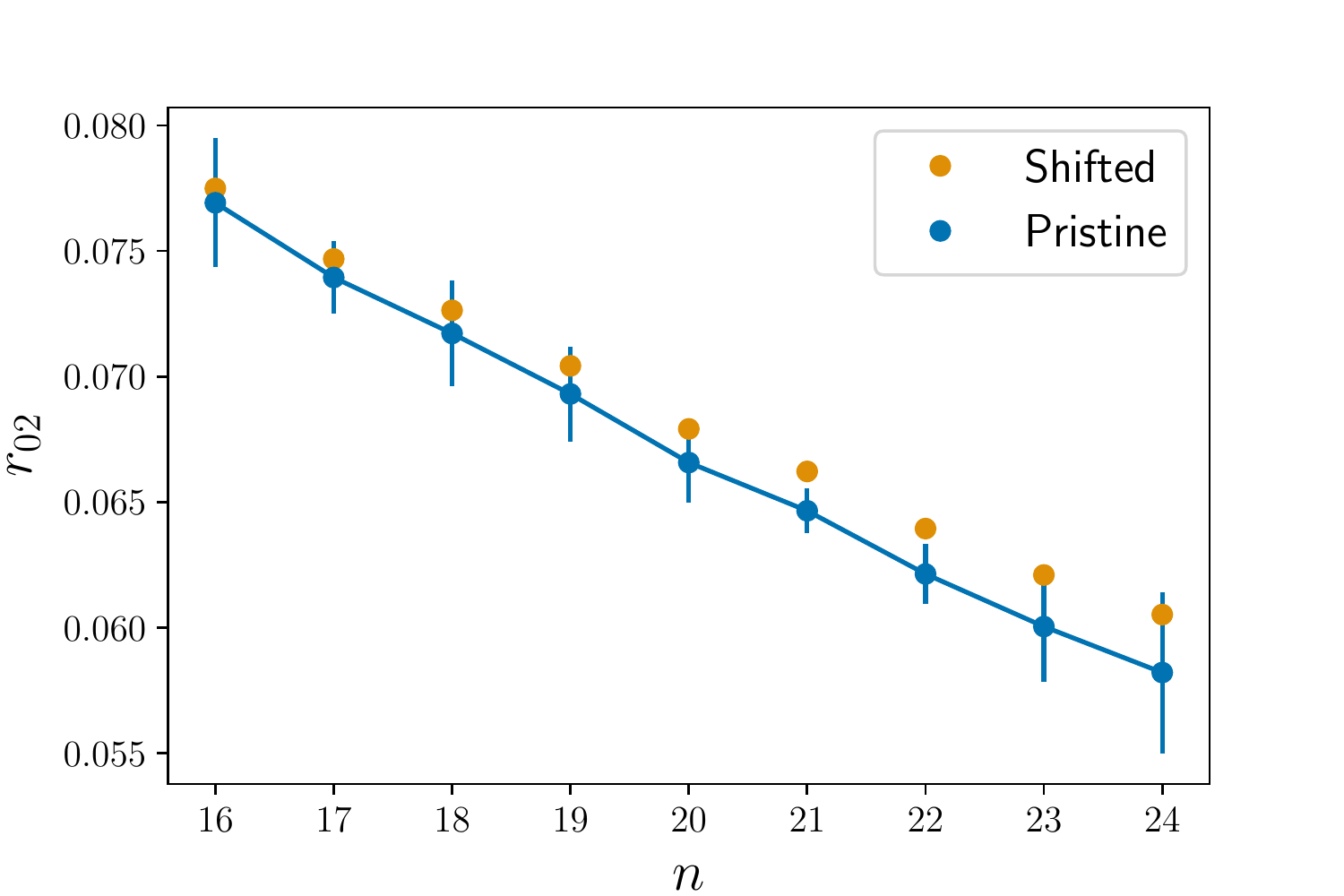}
	\caption{Example of artificially shifted \SR\ with respect to the pristine ratios, calculated using \Brel$=1.2$, $i=0\degree$, \lmin$=11\degree$ and \lmax$=53\degree$. An effect of this size produces an average shifted \SR\ which lies just outside of uncertainty of the average pristine \SR.}
	\label{fig: example shifted r02}
\end{figure}

\begin{figure}
	\centering
	\begin{subfigure}[b]{0.48\textwidth}
		\centering
		\includegraphics[width=\textwidth]{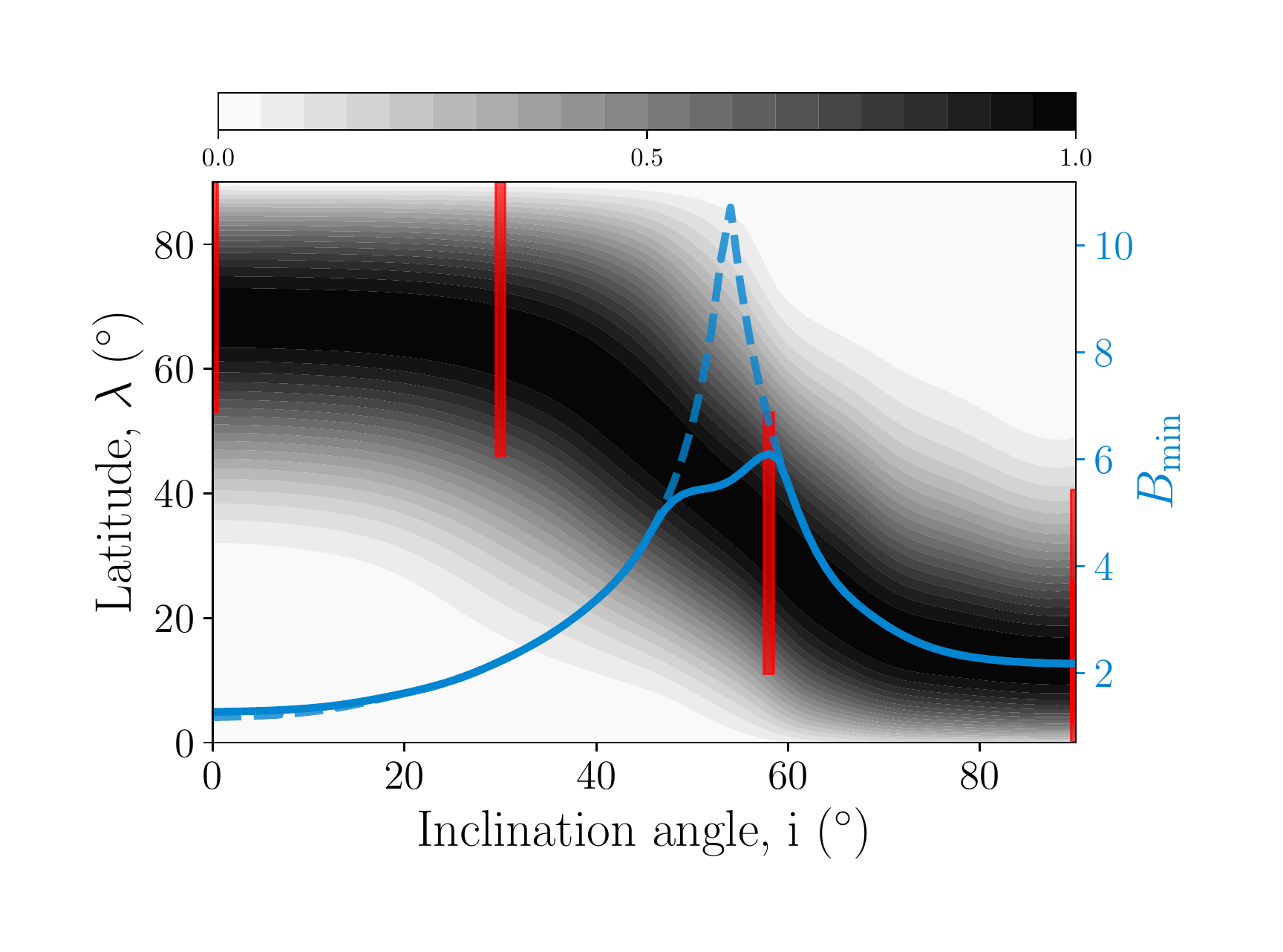}
		\caption{\shift0<\shift2}
		\label{fig: neg combinations}
	\end{subfigure}
	\begin{subfigure}[b]{0.48\textwidth}
		\centering
		\includegraphics[width=\textwidth]{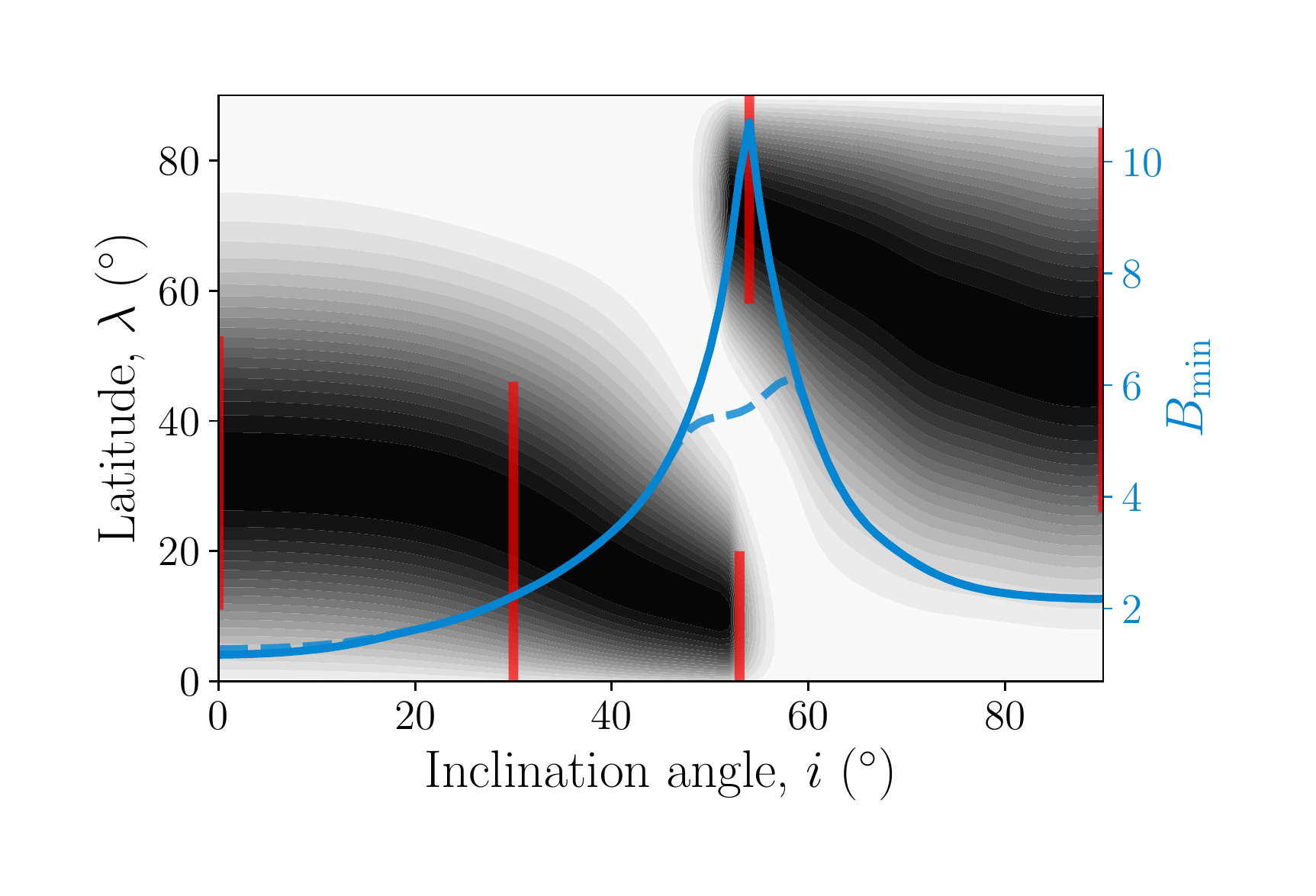}
		\caption{\shift0>\shift2}
		\label{fig: pos combinations}
	\end{subfigure}
	\caption{Determining the combination of parameters to produce separation ratios discernible from the pristine set, where \bmin\ is the minimum field strength needed to do this. The shaded areas show the latitudinal distribution of activity with darker regions indicating where a lower \bmin\ was necessary. The colorbar shows how the shading is inversely proportional to the field strength with the maximum value of 1.0 corresponding to $\frac{1}{B_{\rm min}}$ for that inclination angle. Top: the case for (\shift0)<(\shift2). Bottom: the case for (\shift0)>(\shift2). The solid blue lines are the smallest \bmin\ needed for a particular inclination angle. The dashed blue line is the same but for the opposite sign of shift. The red vertical bars indicate the latitudes occupied by the active band for each set of parameters we chose to focus on (see Table \ref{tab: combinations}). Results were constructed from a grid covering $0<i,\lambda_{\rm min},\lambda_{\rm max}<90\degree$, hence containing $91^3$ models.}
	\label{fig: combinations}
\end{figure}

To choose interesting combinations we first determined those which would produce a set of shifted separation ratios that were, on average, discernibly different from those of the pristine set, i.e beyond the uncertainty of the pristine ratios (Figure \ref{fig: example shifted r02} shows an example set of shifted \SR\ compared to the pristine \SR\ using \Brel$=1.2$, $i=0\degree$, \lmin$=11\degree$ and \lmax$=53\degree$). This was motivated by our goal to find the combinations that would incur a significant bias in stellar property estimates from modelling pipelines. We constructed a grid of $i$, \lmin\ and \lmax , each in the range $0\!-\!90\degree$ with increments of $1\degree$, and calculated the minimum field strength, \bmin, needed to produce the desired shifted separation ratios. \bmin\ was in fact taken to be the weighted average minimum field strength over all of the ratios. Figure \ref{fig: combinations} shows the latitudinal positions of activity bands (shaded regions) used to produce the required separation ratios across the range of inclination angles. For each element in the grid (corresponding to a particular $i$, \lmin\ and \lmax) we shaded the region in latitude and inclination space with a colour intensity that was proportional to $1/$\bmin\ for that element. This was repeated for the entire grid to build up Figure \ref{fig: combinations}. Therefore the darker regions indicate where a lower \bmin\ was necessary to sufficiently shift separation ratios beyond the pristine frequencies, and lighter areas where a much greater \bmin\ was needed.  

For each inclination angle we took the [\lmin,\lmax] pair corresponding to the smallest \bmin\ thereby compressing a 3D grid to 1D. We chose combinations to study both when the \el0 shifts, \shift0, were larger than the \el2 shifts, \shift2, and vice versa. The relative magnitudes of these shifts affect the direction of the bias we get in the stellar property estimations from models. Figure \ref{fig: neg combinations} shows the results for \shift0<\shift2 shifts and Figure \ref{fig: pos combinations} for \shift0>\shift2 shifts.
We can see that for each case the \bmin\ is produced by activity bands at very different latitudes; for example to achieve \bmin\ at high inclinations a lower latitude activity band would produce \shift0<\shift2 whereas a band situated at higher latitudes would cause \shift0>\shift2. 
The solid blue curve in each plot shows the smallest \bmin\ value at each inclination angle and the dashed blue curve shows the same but for the opposite sign of shifts. We can see that at intermediate inclination angles the magnetic field strength would need to be larger than at low or high angles to induce the same size \SR shift. Smaller field strengths are necessary for the lowest and highest inclinations.

A variety of combinations were chosen to generate several sets of shifted frequencies. In addition, we also created artificial data to produce shifts that we would expect from the Sun using $i\!=\!90\degree$ and [\lmin,\lmax]$=[3.3,40.6]\degree$ as found by \cite{2019MNRAS.485.3857T} for the Sun's activity distribution. For one set of frequencies we chose a solar-like magnetic field strength (i.e. \Brel$=\!1$). For another we used the same latitudes and the \bmin\ value required to produce shifted ratios discernibly different from the pristine frequencies (which is higher than we see in the Sun). The red bars in Figure \ref{fig: combinations} correspond to the chosen latitudes and inclination angles. The collections of parameters we used to produce each data set are summarised in Table \ref{tab: combinations}.

\begin{table}
	\centering
	\caption{Combinations of parameters used to calculate frequency shifts for artificial data sets. The method for choosing parameters is explained in the main text. The pristine data set is representative of a field-free star. The Sun model uses the parameters necessary to produce solar-like frequency shifts from our model. The Sun 2 model is the same but for a Sun with stronger magnetic field strength in order to make significantly shifted separation ratios (i.e. \Brel$=$\bmin).}
	\begin{tabular}{ccccc}
		\hline
		\multicolumn{1}{p{0.7cm}}{} & 
		\multicolumn{1}{p{2.1cm}}{Relative magnetic field strength, \Brel} & \multicolumn{1}{p{1.1cm}}{Inclination angle, $i (\degree)$} & \multicolumn{1}{p{1.1cm}}{Minimum latitude, $\lambda_{\rm min} (\degree)$} & \multicolumn{1}{p{1.3cm}}{Maximum latitude, $\lambda_{\rm max} (\degree)$} \\ \hline \hline
		pristine &  0.0  &   -  &   -  &   -   \\ \hline
		i        &  1.2  &   0  &  11  &  53   \\
		ii       &  2.2  &  30  &   0  &  46   \\
		iii      &  9.7  &  53  &   0  &  20   \\
		iv       & 10.7  &  54  &  58  &  90   \\
		v        &  2.2  &  90  &  26  &  85   \\
		vi       &  1.7  &   0  &  53  &  90   \\
		vii      &  2.2  &  30  &  46  &  90   \\
		viii     &  6.1  &  58  &  11  &  53   \\
		ix       &  2.2  &  90  &   0  &  26   \\ \hline
		Sun      &  1.0  &  90  &  3.3 &  40.6 \\
		Sun 2    &  4.0  &  90  &  3.3 &  40.6 \\
		\hline 
	\end{tabular}
	\label{tab: combinations}
\end{table}

We fitted stellar models to our sets of artificial frequencies and compared deviations between their estimations of stellar properties from application to the pristine frequencies and those from the activity-shifted sets. Any differences must be due to the simulated magnetic effects. If activity does not affect stellar model predictions then all results will be similar to those obtained from pristine frequencies.

\section{The modelling pipelines}
\label{sect: stellar models}
We used two different pipelines to fit stellar models to oscillation frequencies. Both used a predefined grid of models. There are many different choices for how to carry out the analysis and the constraints to use when fitting which will impact the uncertainty on estimated stellar properties. In particular, anchoring the lowest frequency modes can reduce error bars as we will show later. Below we detail the input physics and briefly cover the methods of each pipeline, one of which implemented anchoring.

\subsection{\texorpdfstring{\gls{AIMS}}{AIMS}}
The \gls{AIMS} pipeline \citep{2016ascl.soft11014R} uses individual oscillation frequencies, or in this case frequency ratios, along with classical constraints to determine global stellar properties. The grid of models we used was the same as the MS grid from \cite{rendle_aims_2019}. Models were computed using the CL\'{E}S (Code Li\'{e}geois d'\'{E}volution Stellaire, \cite{CLES_evol_code}) stellar evolution code and the grid was parameterised by mass in the range 0.75\msun to 2.25\msun with an interval of 0.02\msun, initial metallicity ($Z_{\rm init}$) from $0.0032-0.0300$ and initial hydrogen content ($X_{\rm init}$) in the range $0.691-0.745$. The $Z_{\rm init}$ and $X_{\rm init}$ values used can be found in Table 1 of \cite{rendle_aims_2019}. Microscopic diffusion with a fixed solar-calibrated mixing length of $1.81$ was included since \cite{rendle_aims_2019} found it to produce more closely matching values for the Sun \citep{thoul}. The convective overshoot was $0.05$ times the local pressure scale height, \cite{1993oee..conf...15G} abundances were used to convert $\rm[Fe/H]$ to $Z/X$, and nuclear reaction rates from taken \cite{2011RvMP...83..195A}. The models were computed using opacities from \cite{1996ApJ...464..943I} and the equation of state from FreeEOS \citep{2012ascl.soft11002I}. Frequencies were calculated using the LOSC (Li\`{e}ge Oscillation Code, \cite{LOSC_puls_code}) pulsation code.

AIMS combines approximating a set of best fitting models using a \gls{mcmc} algorithm (\texttt{emcee}, \cite{2013PASP..125..306F}) with interpolation implemented within the grid of models in order to refine constraints on properties. Interpolation is conducted using multidimensional Delaunay tessellation (see e.g. \cite{field_generic_1991}) both linearly along an evolutionary track and between tracks.

Further details may be found in \cite{rendle_aims_2019}.

\subsection{\texorpdfstring{\gls{YGM}}{YGM}}
For \gls{YGM} analysis we constructed a uniform grid of models for masses in the range 0.95\msun to 1.05\msun\ with a spacing of 0.01\msun. For each mass, models were created with fifteen values of the mixing length parameter spanning $\alpha_{\rm MLT}=1.4$ to 2.625, initial helium abundances spanning from the primordial helium abundance of 0.248 \citep{steigman2010} to 0.30 in steps of 0.01, and initial [Fe/H] in the range -0.30 to +0.30 in steps of 0.01. We use the \citet{gs98} solar mixture to convert [Fe/H] to $Z/X$. The stars were modelled using the Yale Stellar Evolution Code, YREC \citep{2008Ap&SS.316...31D}. For each of the parameters, the models were evolved from the zero-age main sequence to an age of 8 Gyr. Models were output at intermediate ages.

The models were constructed using the Opacity Project (OP) opacities \citep{op} supplemented with low temperature opacities from \citet{2005ApJ...623..585F}. The OPAL equation of state \citep{nayfonov} was used. All nuclear reaction rates are obtained from \citet{1998RvMP...70.1265A}, except for that of the $^{14}N(p,\gamma)^{15}O$ reaction, for which we use the rate of \citet{2004PhLB..591...61F}. All models included gravitational settling of helium and heavy elements using the formulation of \citet{thoul}. The frequencies of the models were  calculated with the code of \citet{antia}.

To determine stellar properties, we defined a goodness of fit for each model in the grid as follows. For each of the spectroscopic observables, [Fe/H], $T_{\rm eff}$ and luminosity $L$, we define a likelihood. For instance, the likelihood for effective temperature was define as
\begin{equation}
	{\mathcal L}(T_{\rm eff})=C\exp(-\frac{1}{2}\chi^2(T_{\rm eff})),
	\label{eqn:tcal}
\end{equation}
with
\begin{equation}
	\chi^2(T_{\rm eff})=\frac{(T^{\rm obs}_{\rm eff}-T^{\rm model}_{\rm eff})^2}{\sigma^2_{ T}},
	\label{eqn:chit}
\end{equation}
where $\sigma_{T}$ is the uncertainty on the effective temperature, and $C$ the constant of normalisation. We define the likelihoods for [Fe/H] and $L$ in a similar manner.

We considered the seismic data using the separation ratio \SR. For this we need to take error correlations into account and thus
\begin{equation}
	\chi^2(r_{02})=({\overline{r}_{02}}^{{\rm obs}}-{\overline{r}_{02}}^{{\rm model}})^T{\mathbf C}^{-1}
	({\overline{r}_{02}}^{{\rm obs}}-{\overline{r}_{02}}^{{\rm model}}),
	\label{eqn:chir02}
\end{equation}
where ${\overline{r}_{02}}^{{\rm obs}}$ is the vector defining the observe \SR, ${\overline{r}_{02}}^{{\rm model}}$ is the vector defining the \SR\ for the model at the observed frequency, and ${\mathbf C}$ is the error-covariance matrix. Thus 
\begin{equation}
	{\mathcal L}(r_{02})=D\exp(-\frac{1}{2}\chi^2(r_{02})),
	\label{eqn:r02like}
\end{equation}
$D$ being the normalisation constant.

The total likelihood is then
\begin{equation}
	{\mathcal L}_{\rm total}={\mathcal L}(r_{02}){\mathcal L}(T_{\rm eff}){\mathcal L}({\rm[Fe/H]}){\mathcal L}(L).
	\label{eqn:total}\end{equation}
The likelihood was normalised by the prior distributions of each property in order to convert to a probability density. The medians of the marginalised likelihoods of the ensemble of models was then used to determine the parameters of the star.

However, the total likelihood defined in Equation \ref{eqn:total} can result in erroneously high likelihood for some models. The surface term is smaller at low frequencies than at high frequencies, but the seismic likelihood function defined above does not take this  into account. Presently, it could be possible to have a model with low $\chi^2(r_{02})$ but where the low frequency modes are badly fit. In order to down weight models for which frequency differences are large, we multiply Equation \ref{eqn:total} with the term
\begin{equation}
	{\mathcal L}_{\rm reg}=E\exp(-\frac{1}{100}\chi^2(\nu_{\rm low})),
	\label{eqn:reg}
\end{equation}
where $\chi^2(\nu_{\rm low})$ is the $\chi^2$ for the two lowest frequency modes of each degree and $E$ is another normalisation constant. Note that Equation \ref{eqn:reg} is not a true likelihood function; the division of the $\chi^2$ by 100 rather than 2 ensures that this term does not dominate the final selection process. One can set this anchoring of the low frequency modes in \gls{AIMS}, however in order to show the range of results to expect using different analysis approaches here we ran \gls{AIMS} without this constraint.


\section{Results}
\label{sect: results}
We determined stellar properties for the pristine and activity-affected stars using the pipelines described in the previous section. Both methods were supplied with the artificial sets of frequencies and fitted using the separation ratios \SR. Observational constraints of effective temperature $T_{\rm eff}=5777\pm80$K and  metallicity $[\rm Fe/H]=0\pm0.1$dex were also provided and were the same for all datasets. Additionally, we refit the same frequency sets, this time with an extra constraint of luminosity $L=1.00\pm0.03$L$_{\odot}$. The assumed use of a luminosity uncertainty of $3\%$ was based on \textit{Gaia} \citep{2018A&A...616A...1G} parallaxes for Sun-like stars. Figures 
\ref{fig: noL, AIMS pristine} and \ref{fig: noL, Sarbani pristine} show results from fitting using just frequencies, $T_{\rm eff}$ and $[\rm Fe/H]$; Figures \ref{fig: withL, AIMS pristine} and \ref{fig: withL, Sarbani pristine} show results from fitting including a luminosity constraint.

Presented are the median results taken from the posteriors of each property, either from the \gls{AIMS} (Panels (a)) or the \gls{YGM} pipeline (Panels (b)). The black circles indicate the median value of stellar properties obtained by fitting to a `pristine', i.e. field-free, set of frequencies which act as a reference. The results obtained from other data sets have been spread along the x-axis for clarity. The grey band shows the uncertainties from $68\%$ confidence intervals on the estimates from stellar models applied to pristine frequencies. The horizontal grey dashed lines illustrate the underlying properties from the corresponding model used to generate pristine frequencies. There is a systematic offset between the underlying properties used to compute the pristine data set and the results from fitting to the pristine frequencies, however they are generally well within error. The focus of this work is on how the results from fitting to shifted frequency ratios differ to those from fitting to pristine frequency ratios since this will be due to magnetic activity effects.

\begin{figure*}
	\centering
	\begin{subfigure}[b]{0.7\textwidth}
		\includegraphics[width=\textwidth]{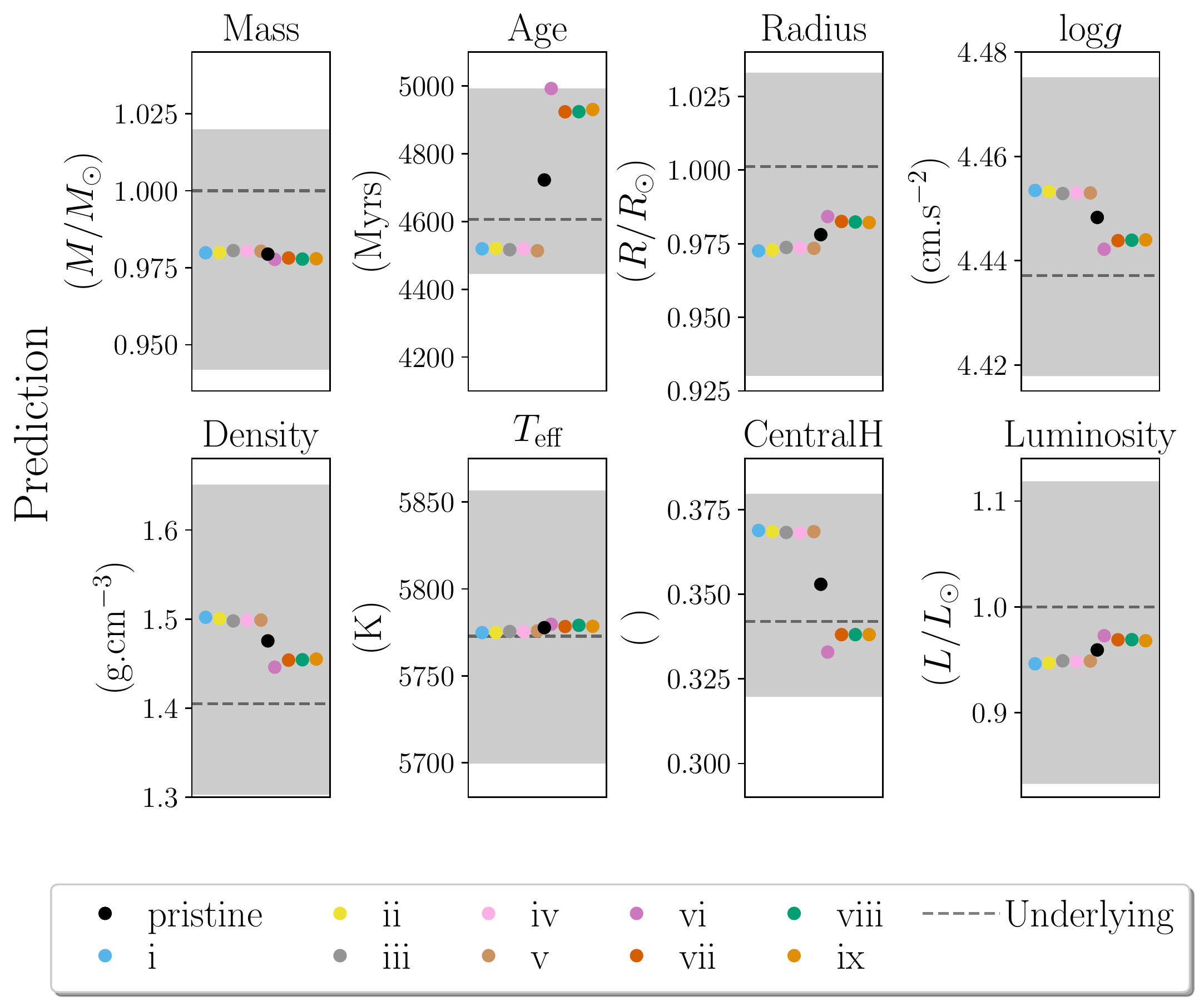}
		\caption{AIMS fitting.}
		\label{fig: noL, AIMS pristine, AIMS fit}
	\end{subfigure}
	\begin{subfigure}[b]{0.7\textwidth}
		\includegraphics[width=\textwidth]{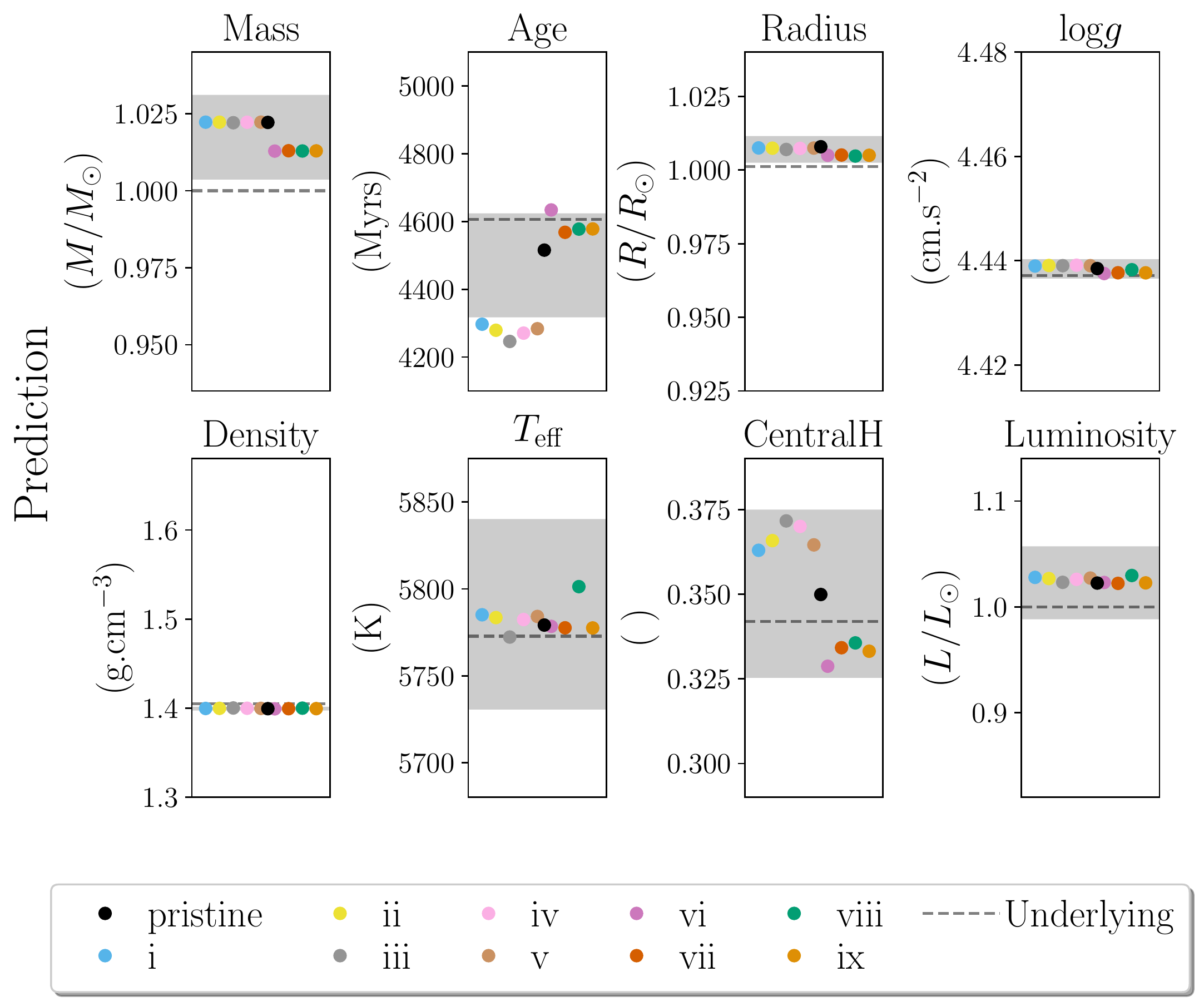}
		\caption{\gls{YGM} fitting.}
		\label{fig: noL, AIMS pristine, new grid fit}
	\end{subfigure}
	\caption{Predictions from fitting without a luminosity constraint. The pristine data set was based on the most solar-like model from the \gls{AIMS} grid. Black circles indicate median results from the pristine data set with the grey band showing the uncertainty on the stellar model estimates from the pristine frequencies. The coloured circles correspond to the results from frequency sets shown in Table \ref{tab: combinations}, and are spread along the x-direction for clarity; their position along the x-axis has no meaning. The same y-axis ranges have been used in Figures \ref{fig: noL, AIMS pristine}-\ref{fig: Sun results} to more easily allow a like-for-like comparison between plots.}
	\label{fig: noL, AIMS pristine}
\end{figure*}

\begin{figure*}
	\centering
	\begin{subfigure}[b]{0.7\textwidth}
		\includegraphics[width=\textwidth]{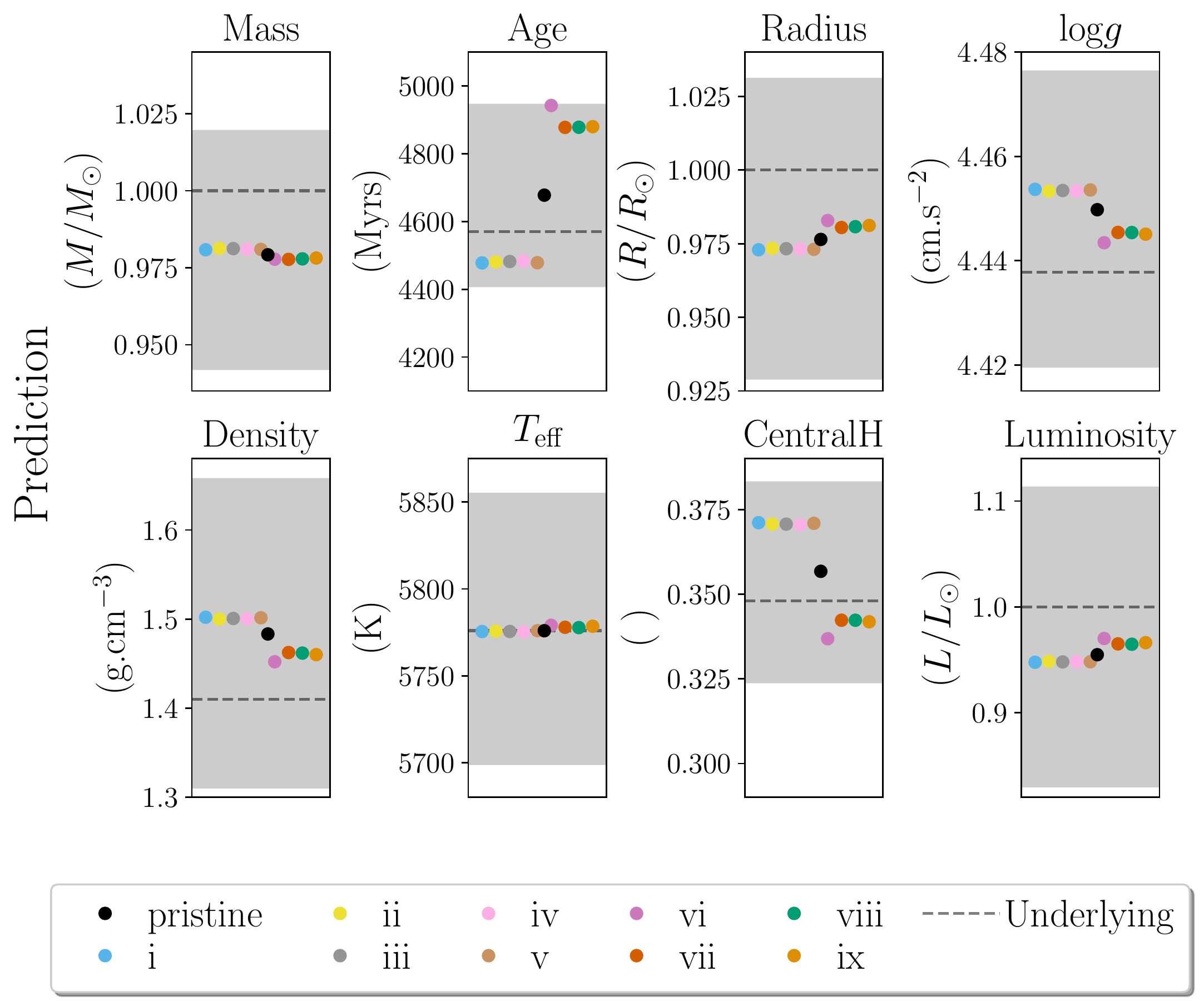}
		\caption{AIMS fitting.}
		\label{fig: noL, Sarbani pristine, AIMS fit}
	\end{subfigure}
	\begin{subfigure}[b]{0.7\textwidth}
		\includegraphics[width=\textwidth]{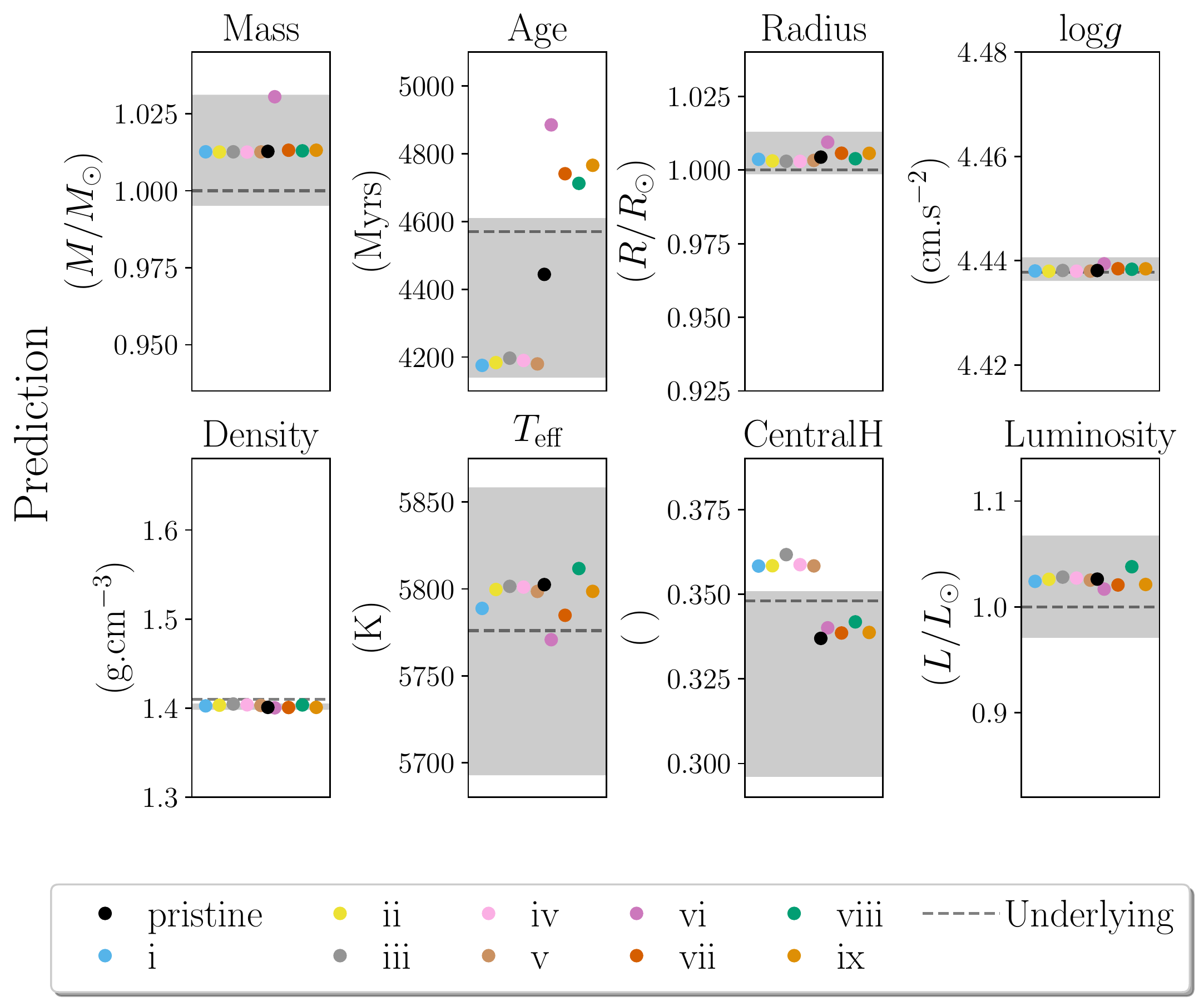}
		\caption{\gls{YGM} fitting.}
		\label{fig: noL, Sarbani pristine, new grid fit}
	\end{subfigure}
	\caption{The same as in Figure \ref{fig: noL, AIMS pristine} but with the pristine data set based on the \gls{ssm}.}
	\label{fig: noL, Sarbani pristine}
\end{figure*}

\begin{figure*}
	\centering
	\begin{subfigure}[b]{0.7\textwidth}
		\includegraphics[width=\textwidth]{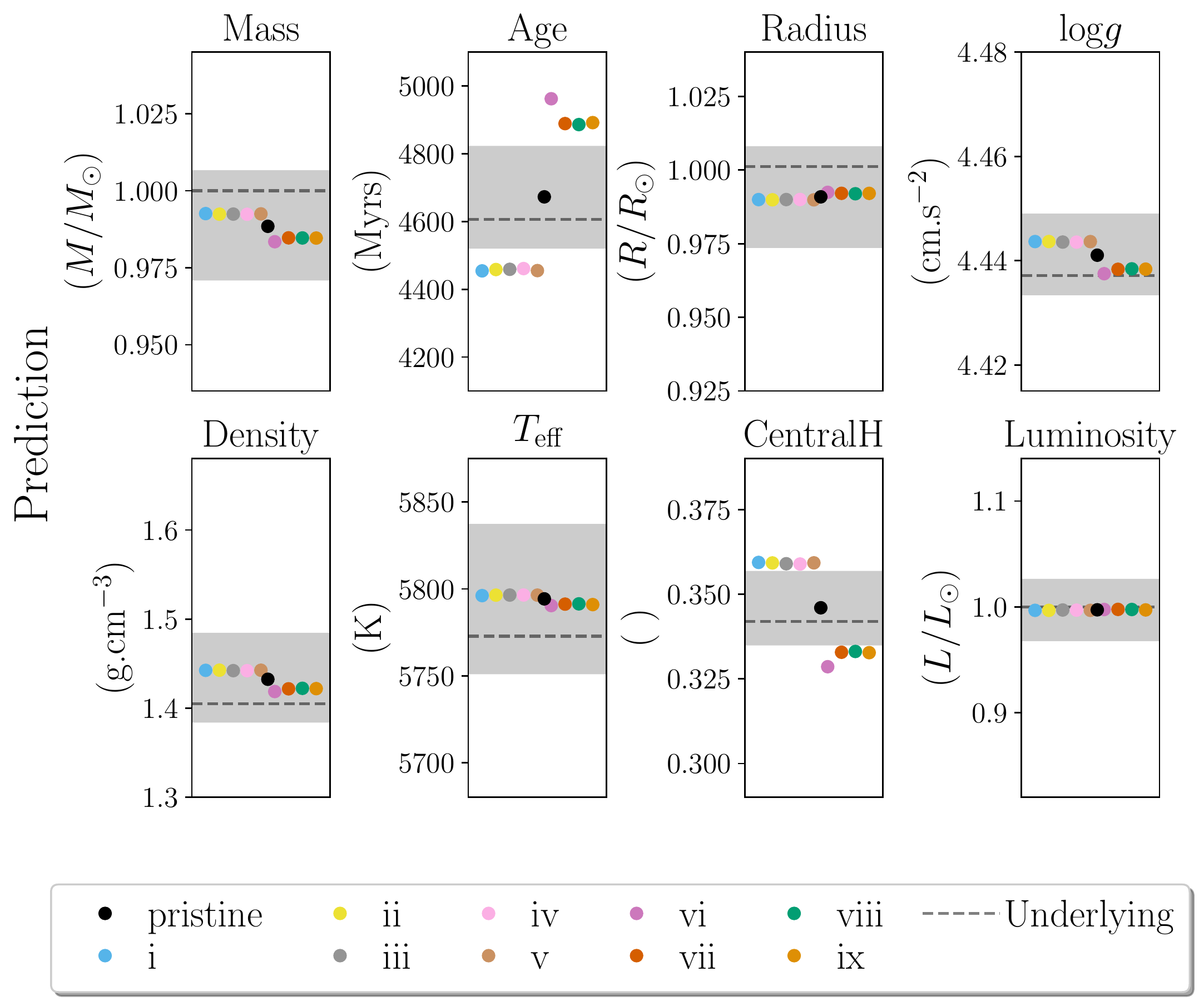}
		\caption{AIMS fitting.}
		\label{fig: withL, AIMS pristine, AIMS fit}
	\end{subfigure}
	\begin{subfigure}[b]{0.7\textwidth}
		\includegraphics[width=\textwidth]{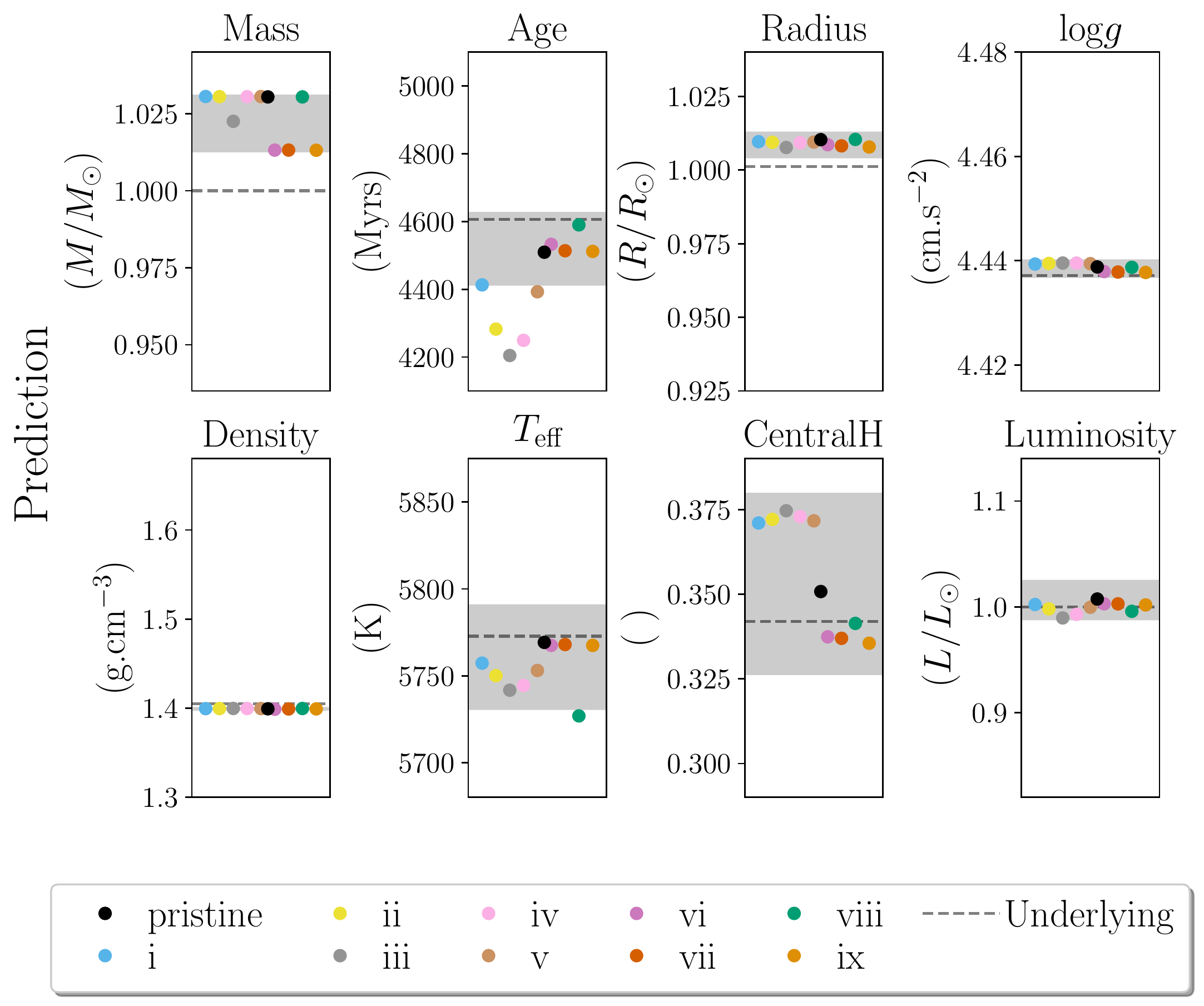}
		\caption{\gls{YGM} fitting.}
		\label{fig: withL, AIMS pristine, new grid fit}
	\end{subfigure}
	\caption{The same as in Figure \ref{fig: noL, AIMS pristine} with the pristine data set based on the most solar-like model from the \gls{AIMS} grid but applying a luminosity constraint.}
	\label{fig: withL, AIMS pristine}
\end{figure*}

\begin{figure*}
	\centering
	\begin{subfigure}[b]{0.7\textwidth}
		\includegraphics[width=\textwidth]{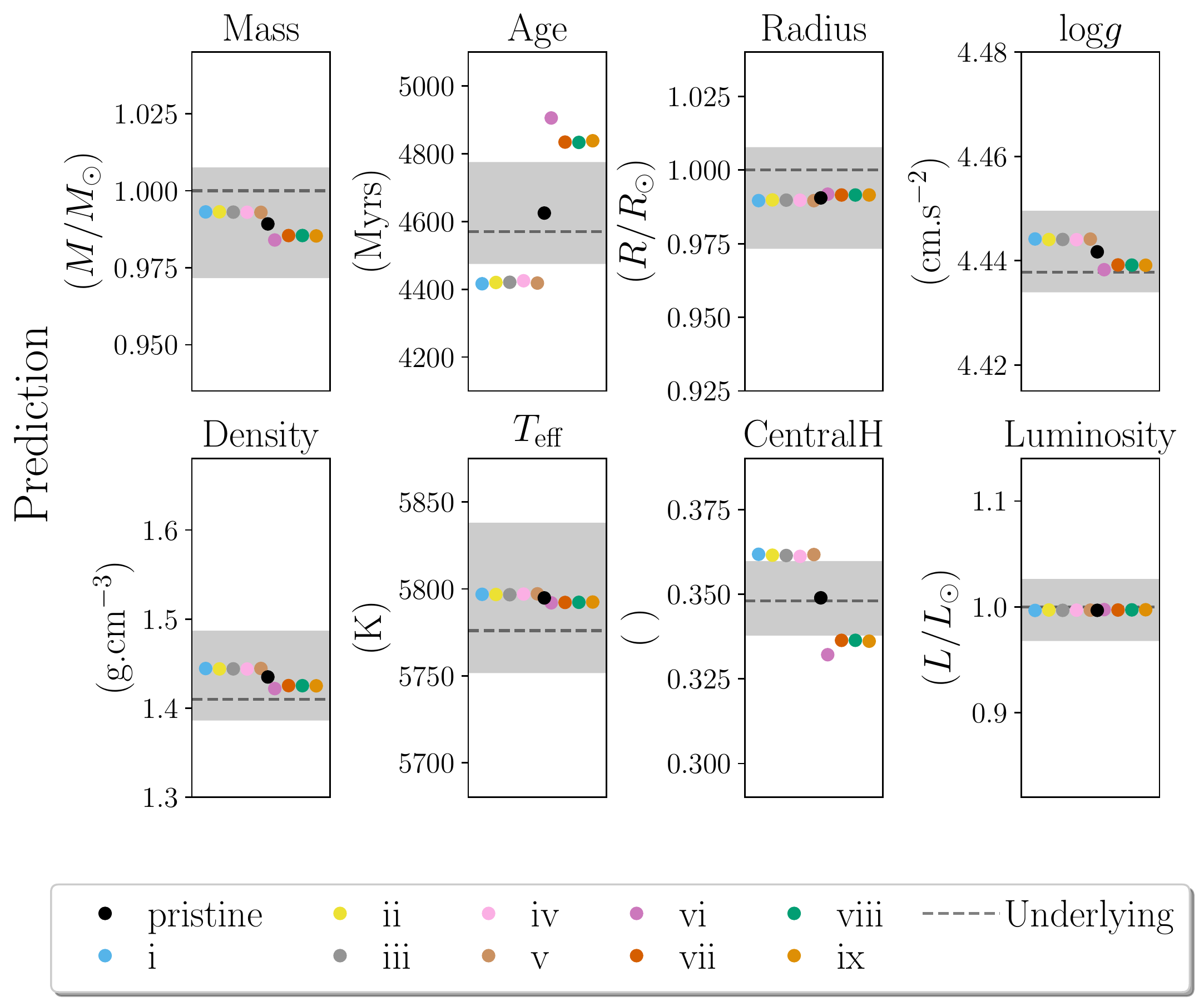}
		\caption{AIMS fitting.}
		\label{fig: withL, Sarbani pristine, AIMS fit}
	\end{subfigure}
	\begin{subfigure}[b]{0.7\textwidth}
		\includegraphics[width=\textwidth]{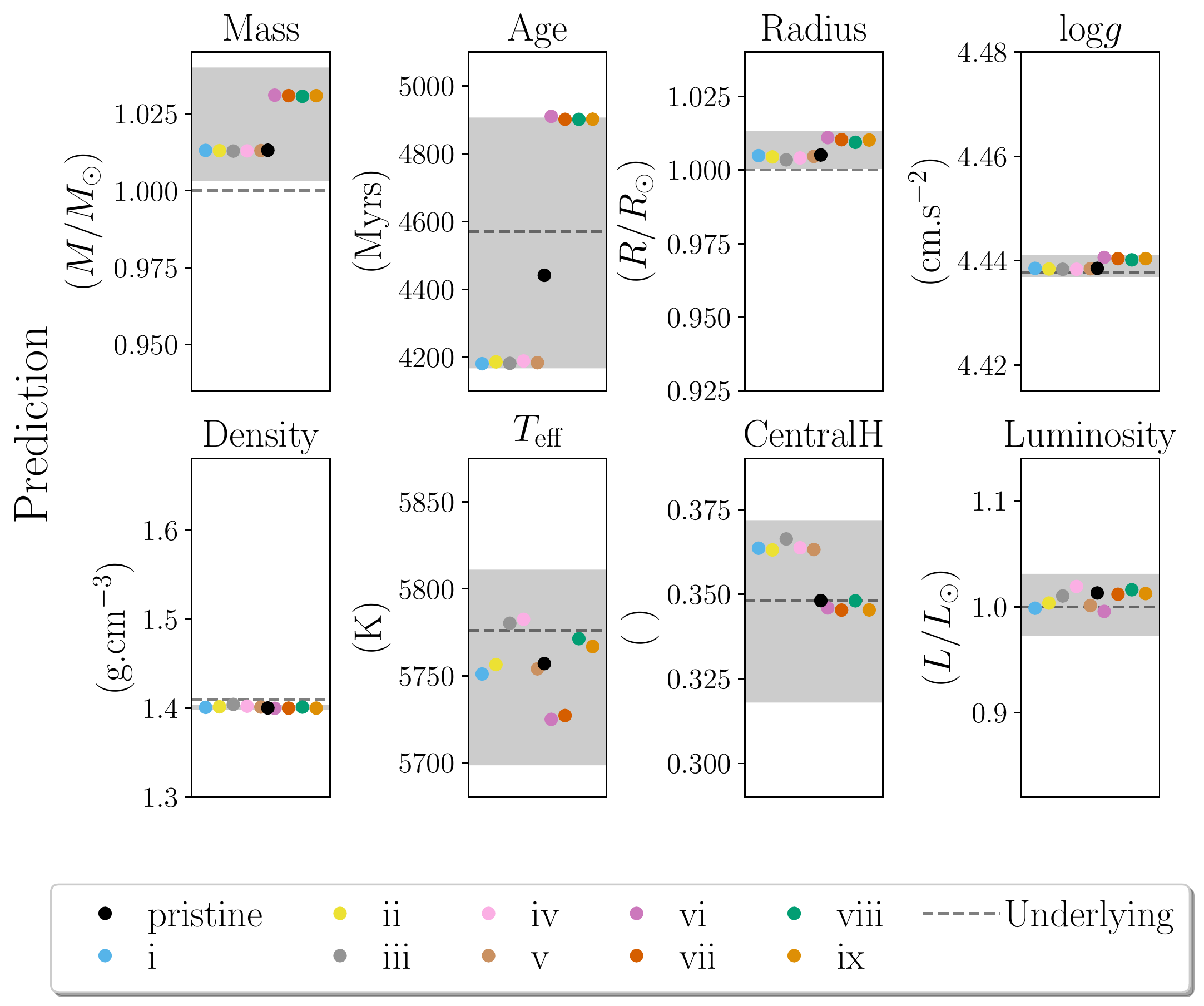}
		\caption{\gls{YGM} fitting.}
		\label{fig: withL, Sarbani pristine, new grid fit}
	\end{subfigure}
	\caption{The same as Figure \ref{fig: noL, AIMS pristine}) but with the pristine data set based on the \gls{ssm} and applying a luminosity constraint.}
	\label{fig: withL, Sarbani pristine}
\end{figure*}


For the majority of properties the median estimates for the activity-affected data lie within the uncertainties of the equivalent pristine values. However, for all runs we can see that the largest differences between estimates from the pristine frequencies and those for different data sets are in age and central hydrogen abundance. For the case of the \gls{YGM} pipeline we can see there is also a considerable spread in the $T_{\rm eff}$ values. The spread is more significant for the fits that included a luminosity constraint since the uncertainty bars are smaller. 

It is clear that for stars experiencing this amount of activity-induced frequency shift, some of the stellar properties we infer will have a notable bias. Focusing on mass and age we find that the bias can be up to 5\% in age, but only up to 0.5\% in mass. This offset is therefore not a concern since  for the analysis carried out in this work, the age parameter typically has a 4.5\% uncertainty and we see a ~2.5\% error on mass.

As described in Section \ref{sect: data}, there are four combinations of results: \gls{AIMS} fitted to \gls{AIMS} model frequencies, \gls{AIMS} applied to frequencies built on the \gls{ssm} data, \gls{YGM} applied to the \gls{AIMS} model frequencies, and finally \gls{YGM} analysis of the \gls{ssm} frequency sets. In general we see similar results for all combinations. The uncertainties from \gls{YGM} fitting are consistently smaller than those from \gls{AIMS} but this can be attributed to the additional constraints placed on the low frequency modes (see Section \ref{sect: stellar models}). \gls{AIMS} uncertainties match what we would expect from \cite{rendle_aims_2019} (Table 3) when fitting using separation ratios and a luminosity constraint. 

In the plots where the absolute spread of stellar property estimates is discernible from the pristine values we can see that there are two distinct groups of points above and below the pristine results. These correspond to where the underlying frequency shift at \el0 is larger than the \el2 shift and hence whether $d_{02}$ is greater or smaller than for the pristine set. For example, if the shifted $d_{02}$ is larger than the pristine $d_{02}$, i.e. \shift0>\shift2, then the fitting will find a smaller age.

To verify that models were equally well fitted to shifted data sets as to the pristine frequency ratios we calculated the ratios of log-likelihoods between fits and found them to be approximately unity. We also tested the analysis methods described above by replacing the \SR\ frequency constraint with $r_{01}$ data, and separately using both \SR\ and $r_{01}$ data simultaneously. The resulting posterior estimates of stellar properties showed very similar patterns to the \SR\ results presented here.

\subsection{The Sun}
In addition to the various combinations tested above, we also studied the bias we would expect for frequency shifts from a Sun-like star compared to the pristine frequencies. As described earlier, shifts were calculated using $i\!=\!90\degree$, \Brel$=\!1$ and $\lambda_{\rm min,max}=[3.3,40.6]\degree$. Another set of frequencies was computed for the same parameter values except using \Brel$=\!4.0$ to imitate a Sun with stronger magnetic field strength.

\begin{figure*}
	\centering
	\includegraphics[width=0.7\textwidth]{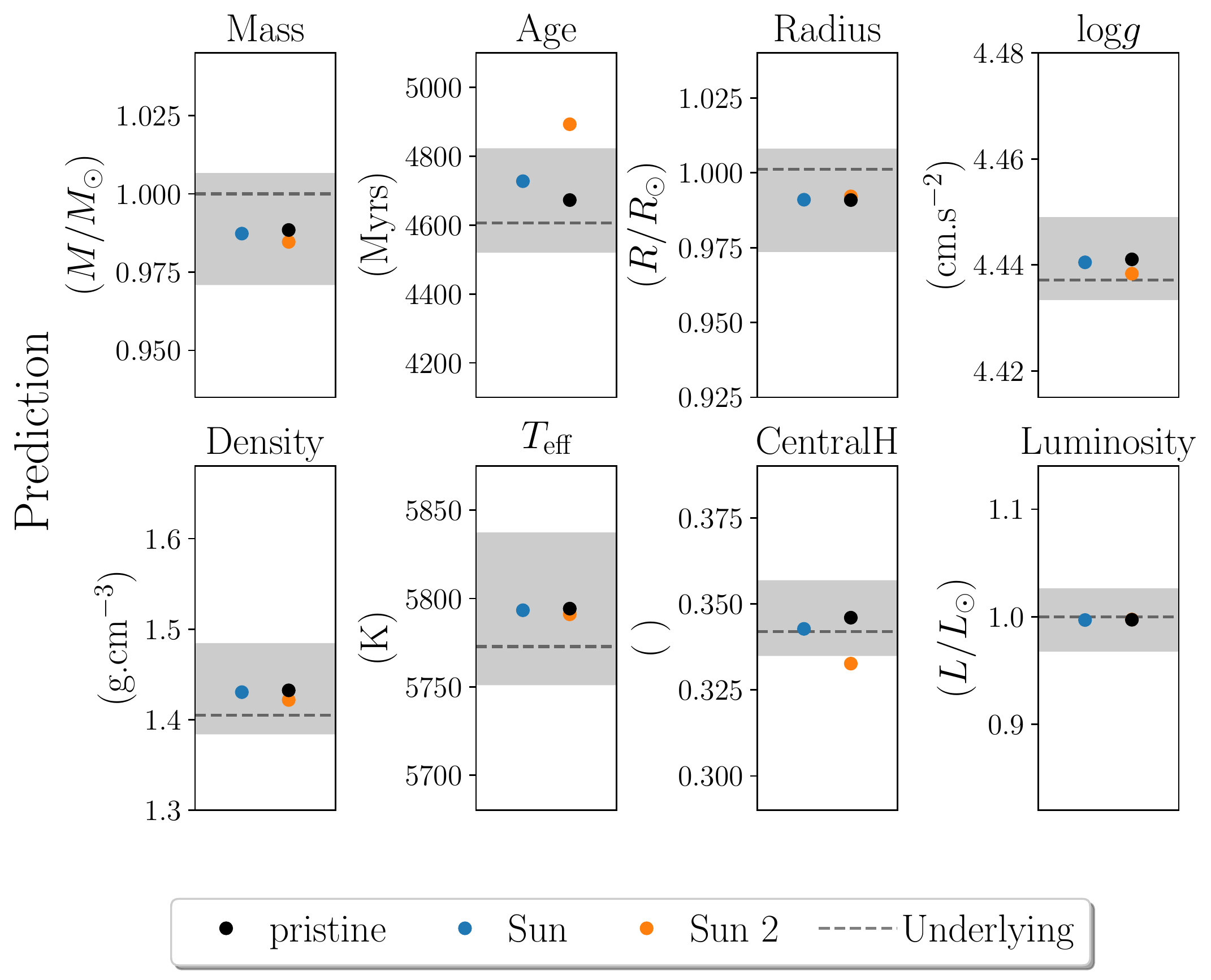}
	\caption{Results from fitting to two additional artificial sets: one using frequency shifts that we would expect to see in the Sun, and the other with same inclination angle and activity distribution as we would observe but with four times the magnetic field strength, i.e. $B=4$. Fitting was conducted by \gls{AIMS} including a luminosity constraint and the pristine data set was based on the most solar-like \gls{AIMS} model.}
	\label{fig: Sun results}
\end{figure*}

Figure \ref{fig: Sun results} shows the results. The estimated stellar properties were not notably different from those one would obtain from pristine frequencies for the solar-like case (\Brel$=\!1$). However, with four-times the field strength we should expect a bias for all properties, being at the 1$\sigma$ level for the estimated age and central hydrogen content.

\section{Discussion and conclusions}
\label{sect: conclusions}

We have demonstrated that for some distributions and strengths of surface stellar activity the oscillation frequencies would experience a shift that impacts the properties obtained from stellar modelling pipelines when applied to separation ratios. The shifts we measure depend on a star's magnetic field strength, the activity distribution on the stellar surface, and it's inclination angle (the angle affecting which azimuthal mode components are detectable). Measured shifts therefore show a complex relationship between these variables. We generated several artificial sets of `measured' frequencies using shifts arising from various combinations of the above. By fitting to separation ratios (\SR) constructed from the frequencies, global properties for these fake stars were estimated by two pipelines and compared to results from a field-free star.

Our results showed that estimates on stellar properties split into two groupings either side of the pristine result based on whether the shifted $d_{02}$ is greater or less than the pristine $d_{02}$. The most noticeable divide is in the age parameter which is lower for an increased $d_{02}$. By extension, given the small range of metallicities here, a lower age will automatically result in a higher central hydrogen abundance. In general we see a greater mass for those shifted data sets with larger $d_{02}$ which is as expected given we are taking a cut in $T_{\rm eff}$.

The division into two groups implies that by measuring the shift in small frequency separation, it is possible to determine the direction of the biases, i.e. whether the property is an under- or over- estimate. Since the size of the bias depends on the frequency shifts experienced by the modes, and is therefore a complex function of the inclination angle, activity strength and distribution, it is more difficult to estimate the size of the bias. By pairing this with the methods of \cite{2019MNRAS.485.3857T} it is possible to constrain the active latitudes present on the star using observations of frequency shifts over time along with the stellar inclination angle. If there is some way to estimate the star's magnetic field strength relative to the Sun then it could be possible to understand the expected size of the bias\textit{} on properties, albeit with fairly large uncertainty. 

We found that, in general, to experience a bias in property estimation larger magnetic field strengths are necessary. The deviations from the underlying properties would be larger for stars with a stronger magnetic field since this simply increases the magnitude of frequency shifts experienced by the modes. For the case of the Sun (\Brel$=\!1$) the frequency shifts due to activity would not produce a significantly biased estimation of solar properties at the levels of precision tested here. In order for these to be affected the field strength would need to be approximately four times stronger. 

The measured frequencies from stars with intermediate inclination angles are least susceptible to magnetic activity effects. As discussed in Section \ref{sect: theory, freq shifts}, this is due to the relative visibilities of the individual azimuthal modes and how their contributions to a central mode frequency are balanced. At these inclinations the field strength would need to be high for the observed separation ratios to be shifted far enough from the pristine \SR\ to have an impact on the estimated stellar properties. This can be seen in Figure \ref{fig: combinations} where the \bmin\ value peaks at $i\!\sim\!55\degree$, where \bmin\ is the field strength required to produce shifted ratios discernibly different to the pristine ratios. However, these inclinations only account for $\sim\!20\%$ of stars (between $45\!-\!60\degree$). Assuming an underlying isotropic distribution of inclination angles, the relative number of stars observed as a function of $i$ is proportional to $\sin(i)$, therefore observing a star with high inclination is more likely. For the lowest ($<\!40\degree$) or highest ($>\!70\degree$) inclinations \bmin\ is lower. This means that for stars at these inclination angles more care must be taken to consider the bias on stellar properties due to activity.
	
For this analysis we took frequency uncertainties commensurate with \Kepler\ data of a year or more. For shorter duration observations the frequency resolution will be reduced thereby minimising these effects since estimates of stellar properties will have larger uncertainties. The significance of the bias in properties will depend on the quality of data provided to the modelling pipeline. This includes additional spectroscopic measurements and whether low-frequency modes are constrained separately to separation ratios.
	
We find that for asteroseismic observations of Sun-like targets we can expect magnetic activity to affect mode frequencies which will bias the results from stellar modelling analysis. For most stellar properties we studied this offset should not be an issue since it is smaller than the uncertainties, including those on mass. However, for age and central hydrogen content the effect could be significant. Particular care must be taken when analysing long duration observations of stars with stronger magnetic field strengths than the Sun for which we expect higher magnitude frequency shifts. The same is true for stars with very high or low inclination angles where, for the same field strength, the shift in measured frequency separation ratios is easier to observe and therefore will produce a more significant bias.

An obvious next step is to assess the fraction of asteroseismic targets in the \Kepler\ and \tess\ samples that might be susceptible to these effects based on results from asteroseismic signatures of stellar activity cycles \citep[e.g.][]{2011A&A...530A.127S,2016A&A...589A.103R,2016A&A...589A.118S,2017A&A...598A..77K,2018ApJS..237...17S,2019FrASS...6...52K} and proxies of magnetic activity (e.g. see \citealt{2019FrASS...6...46M} and references therein).

\section*{Acknowledgements}
We would like to thank Josefina Montelb\'{a}n for her useful discussions regarding the use of \gls{AIMS}. A.E.L.T., W.J.C. and G.R.D. acknowledge the support of the Science and Technology Facilities Council (STFC). Funding for the Stellar Astrophysics Centre is provided by The Danish National Research Foundation (Grant agreement no.:DNRF106). A.M. acknowledges support from the ERC Consolidator Grant funding scheme (project ASTEROCHRONOMETRY, \url{https://www.asterochronometry.eu}, G.A. n. 772293).

\section*{Data Availability}
The data underlying this article were generated with publicly available software: \gls{AIMS}, \url{https://gitlab.com/sasp/aims}.

The data underlying this article will be shared on reasonable request to the corresponding author.

\bibliographystyle{mnras}
\bibliography{SR_magnetic_activity} 

\bsp	
\label{lastpage}
\end{document}